\documentstyle[psfig,preprint,aps,pre]{revtex}

\newcommand{\bm}[1]{\mbox{\boldmath $#1$}}

\begin{document}

%\preprint{draft \today}

\title{Theory for Diffusion-Limited Oscillating Chemical Reactions}
\author{Harmen J. Bussemaker}
\address{Institute for Physical Science and Technology\\
	University of Maryland\\ 
	College Park, Maryland 20742}
\author{Ricardo Brito}
\address{Facultad de Ciencias F\'\i sicas\\
	 Universidad Complutense\\28040 Madrid, Spain\\[6mm]
	\rm (J. Stat. Phys. 87, 1165--1178 (1997).)}

\tighten

\maketitle

% \begin{center}\today\end{center}

\begin{abstract}
A kinetic description of lattice-gas automaton models for reaction-diffusion
systems is presented. It provides corrections to the mean-field rate 
equations in the diffusion-limited regime.
When applied to the two-species Maginu model, the theory 
gives an excellent quantitative prediction of the effect of slow diffusion
on the periodic oscillations of the average concentrations in 
a spatially homogeneous state.\\[12pt]
\end{abstract}

\newpage 

\section{Introduction}

In this paper we describe how a relatively simple theory quantitatively explains
the deviations from mean-field behavior that occur in diffusion-limited
chemical reactions.
The modeling of chemical reactions in spatially extended systems is
an interesting application of a class of microscopic models called 
``lattice-gas automata'' \cite{Doolen}.  
Space, velocity, and time are all discrete in such models, which 
simplifies implementation on computers as well as theoretical analysis.
Lattice-gas automata (LGA) provide a flexible tool for studying the 
various phenomena resulting from the interplay between reaction 
and diffusion~\cite{review-LGA}.

Here we will not be concerned with chemical pattern formation, but instead we
will consider a spatially extended two-species model exhibiting coupled 
periodic 
oscillations of the concentrations of both species in a spatially homogeneous 
state.
If the reactions are slow compared to the diffusion, then mean-field or
Boltzmann theory equations give an excellent description of the reaction 
kinetics.  This is the so-called reaction-limited regime.
In the opposite diffusion-limited case however, when the diffusion is
slow compared to the reactions, there is no time to equilibrate after a 
reaction before another reaction occurs.  
Consequently, equal-time correlations will be present that invalidate the
{\em Stosszahlansatz} or molecular chaos assumption used to derive the 
Boltzmann equation.
Therefore in the diffusion-limited regime the behavior of the system is
seriously modified.

A condition that guarantees the absence of correlations in the equilibrium
state of lattice-gas automata is the so-called detailed balance (DB) condition.
Reactive LGA's in the diffusion-limited regime violate DB.
A systematic theory for LGA's violating DB has recently been developed by 
Ernst and coworkers \cite{Bussemaker95a,Ernst+co}.
In the present paper we apply this theory to calculate corrections to the 
Boltzmann equation.
A similar method has been developed by Boghosian and Taylor 
\cite{Bogho,Bogho-schloegl}.

The organization of this paper is as follows.
In section~\ref{sec:model} we define the model used.
We present the ring kinetic theory in section~\ref{sec:theory},
and compare it with computer simulations in section~\ref{sec:compare}.
We end with a discussion in section~\ref{sec:discussion}.

\section{The Model}\label{sec:model}

\subsection{Reactive Lattice-Gas Automaton}\label{sec:LGA}

In a lattice gas automaton particles live on a regular lattice,
${\cal{ L}}$, so that their positions can only take a a limited set of
values corresponding to the nodes of the lattice. The velocities are also 
restricted, and must be equal to unit vectors oriented along the 
the links connecting the neighboring nodes. We denote this set by 
$\{{\bf c}_i;\ 1\leq i\leq b \}$ where $b$ is the coordination number 
of the lattice. The square lattice, with $b=4$, will be used in this paper
as it has sufficient symmetry to properly describe the diffusive 
problem that we are considering. We further impose an exclusion 
principle requiring that no more that one particle can be at the same node
with the same velocity. As a consequence there can be at most $b$ particles 
per node, i.e., one per link. The state of the LGA is fully described 
by a set of boolean occupation numbers 
$\{s_i({\bf r});\ 1\leq i\leq b,\ {\bf r}\in{\cal L}\}$, 
where $s_i({\bf r})$ equals 1 if there is a particle at node $\bf r$
with velocity ${\bf c}_i$ and 0 otherwise. 

For multi-species models with $m$ types of reactants, such as the Maginu
model where $m=2$, we have to introduce different types of particles.
The exclusion principle has to be modified
in order to allow for the coexistence of several species.
We adopt the coupled-lattice model described in \cite{Dab}.
In this approach particles of different types live on separate lattices,
and only interact when a chemical reaction occurs. The exclusion
principle is applied independently to each lattice. However,
for the sake of compactness in the mathematical derivations, we can extend 
the former set of occupation numbers $s_i({\bf r})$ to a new set
$\{s_i({\bf r});\ 1\leq i\leq mb,\ {\bf r}\in{\cal L} \}$,
in such a way that channels $1\leq i\leq b$ are reserved for particles 
of species 1, channels $b+1\leq i \leq 2b$ for species 2, etc.
The number of particles of type $p$ is given by 
\begin{equation}\label{number}
        \alpha_p({\bf r}) = \sum_{i=1+(p-1)b}^{pb} s_i({\bf r}).
\end{equation} 

A time evolution step is the composition of two substeps, defined as 
follows. 
First, at each node independently a reactive collision takes place,
during which a pre-reaction state 
$s({\bf r}) \equiv \{s_i({\bf r}),\ 1\leq i \leq mb\}$ is replaced by a 
post-reaction state $\sigma({\bf r})$ in a stochastic process governed 
by a set of transition probabilities $A_{s\sigma}$. 
The reactive collision is followed by a propagation step, during which all 
particles are moved to neighboring nodes ${\bf r}+{\bf c}_i$
in the direction of their velocities.

Let us describe the reactive collision step in detail.  The chemical
reaction we want to simulate is described by
\begin{equation}\label{chemist}
	\alpha_1 X_1 + \alpha_2 X_2 +\ldots+ \alpha_m X_m \longrightarrow 
	\beta _1 X_1 + \beta _2 X_2 +\ldots+ \beta_m X_m 
\end{equation}
and occurs at a rate $P(\bm{\alpha},\bm{\beta})$, where
$\bm{\alpha}=(\alpha_1,\alpha_2,\dots)$ and 
$\bm{\beta }=(\beta _1,\beta _2,\dots)$ specify the number of particles
before and after reaction, and $X_p$ represents  species $p$. 
The  outcome of the chemical reaction only depends on the number of 
particles of each species, $\{\alpha_p(s);\ 1\leq p\leq m\}$, 
present at the node before  the reaction, not on the velocity distribution.
After the reaction, the $\beta_p$ particles of each species are randomly
redistributed over the $b$ available velocity directions
(this random redistribution models the diffusion process), 
which can be done in $b!/(\beta_p)!(b-\beta_p)!$ ways for species $p$.
Thus, the transition probability from precollision state $s$ to postcollision
state $\sigma$ is given by
\begin{equation}
  A_{s\sigma} = \left[\prod_{p=1}^m
        \frac{(\beta_p(\sigma))! (b-\beta_p(\sigma))!}{b!} \right]
        P(\bm{\alpha},\bm{\beta}).
\end{equation}
Note that the normalization $\sum_\sigma A_{s\sigma}=1$ follows from
the normalization $\sum_{\bf\beta} P(\bm{\alpha},\bm{\beta})=1$.

\subsection{Maginu Model}\label{sec:Maginu}

The Maginu model \cite{Maginu} is a two species model that exhibits a 
variety of behavior.
It is described by the following equations for the concentrations $x$ and 
$y$ of the two  species  \cite{review-LGA}:
\begin{eqnarray}\label{Maginu}
	\frac{\partial x}{\partial t} &=& x-x^3/3 -y +D_{x} \nabla^2 x
	\nonumber \\  
	\frac{\partial y}{\partial t} &=& (x -ky)/c +D_{y} \nabla^2 y
\end{eqnarray}
with $c>0$ and $0<k<1$. The constants $D_{x}$ and $D_{y}$ are the 
diffusion coefficients for the two  species respectively. 
Depending on the parameters, the model can exhibit Turing structures 
(when $D_{x}$ is very different from $D_{y}$) as well as periodic 
behavior. Here we will solely be interested in the case 
$D_{x}=D_{y}$, where the system 
develops a stable limit cycle in a homogeneous state. 
This limit cycle shrinks as the chemical reaction rate increases.

The Maginu model as defined by Eq.~(\ref{Maginu}) is not directly useful
since the concentrations can become negative,
and therefore cannot be simulated with an LGA \cite{Dab}. 
This problem is however easily solved by using the linear transformation
\begin{eqnarray}\label{transformation} 
	x &=& \case{1}{2}+x/\sqrt{12(1+k)/k},   \nonumber  \\ 
	y &=& \case{1}{2}+yk/\sqrt{12(1+k)/k}
\end{eqnarray} 
where $x$ and $y$ are the concentrations of the two species 
$X$ and $Y$   that we will study. 

Next we have to determine a set of reaction rates  
$ P(\bm{\alpha},\bm{\beta})$ for the LGA that gives rise to the macroscopic
behavior defined by Eqs.~(\ref{Maginu}) and (\ref{transformation}). 
The matrix $P(\bm{\alpha},\bm{\beta})$ is needed
in numerical simulations as well as in the theory presented in
the next section. 
In Ref.~\cite{review-LGA} a method for constructing
$P(\bm{\alpha},\bm{\beta})$ has been extensively discussed,
and we will not give the details here. We will however, adopt 
the rules of Ref.~\cite{Dab}, where the number of particles is only
allowed to change by $\pm 1$ during the reaction. 
The matrix $P(\bm{\alpha},\bm{\beta})$ is then uniquely specified. 

An important point in the definition of the collision rules is the 
introduction of a {\em time scaling parameter}, $s$, which allows us
to control whether the system is in the reaction-limited or in the
diffusion-limited regime (see Ref.~\cite{Dab} for details).
For large values of $s$ we have 
$P(\bm{\alpha},\bm{\beta}) \simeq\delta (\bm{\alpha},\bm{\beta})$ 
(where $\delta$ is the Kronecker delta):
chemical reactions occur at a very slow rate.
This is the diffusion-limited regime,
where diffusion is able to maintain the homogeneity in the system, 
and where Eqs.~(\ref{Maginu}) and (\ref{transformation}) are meaningful, 
as the conditions for their derivation are fulfilled. 
On the other hand, for small values of $s$ chemical reactions occur at a
much faster rate, and diffusion is no longer able to maintain spatial
homogeneity. This is the reaction-limited regime.
In the next section we present a theory that explains the
behavior of the system throughout both regimes.

\section{Ring kinetic theory}\label{sec:theory}

In mean-field or Boltzmann approximation all correlations between
occupation numbers are neglected, and the state of the system is completely
specified by the average occupation numbers,
\begin{equation}
	f_i({\bf r},t)=\langle s_i({\bf r},t) \rangle.
\end{equation}
The time evolution of $f_i({\bf r},t)$ is given by the nonlinear Boltzmann
equation,
\begin{equation}\label{f-i-r-t}
	f_i({\bf r},t+1) = f_i({\bf r},t) + I_i[f({\bf r},t)].
\end{equation}
The nonlinear collision operator is defined as
\begin{equation}
	I_i[f] = \sum_{s,\sigma} (\sigma_i-s_i) A_{s\sigma} F(s)
		 \equiv \langle \sigma_i-s_i \rangle_F.
\end{equation}
We have introduced $\langle\cdots\rangle_F$ as an average that 
assumes that the precollision state is factorized over all channels,
so that the probability to find a state $s$ is given by
\begin{equation}\label{factor-F}
	F(s) = \prod_i f_i^{s_i} (1-f_i)^{1-s_i}.
\end{equation}
In this approximation, where $F(s)$ is given by Eq.~(\ref{factor-F}),
and the transition rates $A_{s\sigma}$ are those of the Maginu model, 
the nonlinear Boltzmann equation (\ref{f-i-r-t}) is equivalent to the 
mean-field rate equations (\ref{Maginu}) and (\ref{transformation}).

To go beyond the mean-field approximation we consider the pair correlation 
function,
\begin{equation}\label{Cdef}
	{\cal C}_{ij}({\bf r}-{\bf r}',t) 
	= \langle \delta s_i({\bf r},t) \delta s_j({\bf r}',t) \rangle.
\end{equation}
Here we have assumed that the system is translationally invariant.
The fluctuations are defined as $\delta s_i=s_i-f_i$.
A special role is played by the on-node correlations ${\cal C}_{ij}({\bf 0},t)$;
by definition the diagonal elements vanish: 
${\cal C}_{ii}({\bf 0},t) \equiv 0$. 
We neglect all triplet and higher order correlations.
In a spatially homogeneous system, where $f_i({\bf r},t)=f_i(t)$,
the time evolution of $f_i(t)$ is then described by the generalized 
Boltzmann equation,
\begin{equation}\label{gbeq}
	f_i(t+1) - f_i(t) = I_i[f(t)] 
		+ \sum_{k<l}I^\prime_{i,kl}[f(t)] {\cal C}_{kl}(t).
\end{equation}
Here the operator $I^\prime$ describes corrections to the Boltzmann collision 
term $I$.  It is defined by
\begin{equation}
	I^\prime_{i,kl}[f] 
	= \frac{\partial^2 I_i[f]} {\partial f_k \partial f_l}
	= \left\langle (\sigma_i-s_i) \frac{\delta s_k \delta s_l}{g_k g_l} 
	  \right\rangle_F,
\end{equation}
where $g_i=\langle (\delta s_i)^2\rangle=f_i(1-f_i)$ is the single channel
fluctuation.

In order to have a complete theory we must provide a time evolution 
equation for ${\cal C}_{ij}({\bf r},t)$.
To derive this equation we will make the important assumption that
the average occupations {\em change slowly in time}.
In fact, as far as the evolution of ${\cal C}_{ij}({\bf r},t)$ is concerned,
we will assume that no chemical reactions occur at all, so that the model is
purely diffusive.
Under this assumption, the average occupations in equilibrium are given by
\begin{eqnarray}
	f^{\rm eq}_i = x &\qquad& (i=1,2,3,4) \nonumber\\
	f^{\rm eq}_i = y &\qquad& (i=5,6,7,8),
\end{eqnarray}
where $x$ and $y$ are the average concentrations of species $X$ and $Y$,
respectively.
When $f_i({\bf r},t)$ is close to equilibrium, the approach to equilibrium
is given by the linearized Boltzmann equation
($\delta f_i = f_i - f^{\rm eq}_i$),
\begin{equation}
	\delta f_i({\bf r}+{\bf c_i},t+1) 
	= \sum_j (\openone+\Omega)_{ij} \delta f_j({\bf r},t),
\end{equation}
where $\openone_{ij}=\delta_{ij}$ and
the linearized Boltzmann operator is defined by
\begin{equation}
	\Omega_{ij} = \frac{\partial I_i[f]} {\partial f_j}
	= \left\langle (\sigma_i-s_i) \frac{\delta s_j}{g_j} \right\rangle_F.
\end{equation}
Under the assumption of slow reactions we have
\begin{equation}
	(\openone+\Omega)
	= \frac{1}{4}\left(\begin{array}{llllllll}
	%	1/4 & 1/4 & 1/4 & 1/4 &
		1\  & 1\  & 1\  & 1\  &
		0\  & 0\  & 0\  & 0  \\
	%	1/4 & 1/4 & 1/4 & 1/4 &
		1 & 1 & 1 & 1 &
		0 & 0 & 0 & 0 \\
	%	1/4 & 1/4 & 1/4 & 1/4 &
		1 & 1 & 1 & 1 &
		0 & 0 & 0 & 0 \\
	%	1/4 & 1/4 & 1/4 & 1/4 &
		1 & 1 & 1 & 1 &
		0 & 0 & 0 & 0 \\
		0 & 0 & 0 & 0 &
	%	1/4 & 1/4 & 1/4 & 1/4 \\
		1 & 1 & 1 & 1 \\
		0 & 0 & 0 & 0 &
	%	1/4 & 1/4 & 1/4 & 1/4 \\
		1 & 1 & 1 & 1 \\
		0 & 0 & 0 & 0 &
	%	1/4 & 1/4 & 1/4 & 1/4 \\
		1 & 1 & 1 & 1 \\
		0 & 0 & 0 & 0 &
	%	1/4 & 1/4 & 1/4 & 1/4 \\
		1 & 1 & 1 & 1 \\
	\end{array} \right).
\end{equation}
It is natural to assume that fluctuations $\delta s_i({\bf r},t)$ will 
decay to equilibrium in a manner similar to $\delta f_i({\bf r},t)$, i.e.,
\begin{equation}\label{delta_s}
	\delta s_i({\bf r}+{\bf c}_i,t+1) 
	= \sum_j (\openone+\Omega)_{ij} \delta s_j({\bf r},t).
\end{equation}
However, two fluctuations at the {\em same} node will be correlated after
collision, even if before collision the distribution is completely factorized.
This is a consequence of the violation of detailed balance \cite{Bussemaker95a}.
The generation of on-node postcollision correlations is quantified by
\begin{equation}\label{OM20}
	\Omega^{2,0}_{ij}[f(t)] = \langle \delta\sigma_i({\bf r},t) 
		\delta\sigma_j({\bf r},t) \rangle_F.
\end{equation}
This expression vanishes in the non-reactive limit $s\to\infty$.
The presence of on-node correlations ${\cal C}_{ij}({\bf 0},t)$ before 
collision gives rise to corrections to $\Omega^{2,0}_{ij}[f(t)]$,
and the full postcollision source term is given by 
(see Ref.~\cite{Bussemaker95a})
\begin{equation}\label{source}
	B_{ij}(t) = \Omega^{2,0}_{ij}[f(t)] + {\cal C}_{ij}({\bf 0},t)
		+ \sum_{k,l}\Omega^{2,2}_{ij,kl}[f(t)] {\cal C}_{kl}({\bf 0},t),
\end{equation}
where $\Omega^{2,2}_{ij,kl}[f] = \partial^2 \Omega^{2,0}_{ij}[f]/
\partial f_k \partial f_l$.
Combining Eqs.~(\ref{delta_s}) and (\ref{source}) with the definition of
${\cal C}_{ij}({\bf r},t)$ in Eq.~(\ref{Cdef}) we obtain the ring kinetic 
equation
\begin{equation}\label{ring-equation}
	{\cal C}_{ij}({\bf r}+{\bf c}_i-{\bf c}_j,t+1)
	= (1-\delta_{{\bf r,0}}) \sum_{k,l} (\openone+\Omega)_{ik} 
	  (\openone+\Omega)_{jl} {\cal C}_{kl}({\bf r},t)
	  + \delta_{{\bf r,0}} B_{ij}[f(t)].
\end{equation}
This equation has been derived in a more systematic fashion in 
Ref.~\cite{Bussemaker95a}.

The physical interpretation of Eq.~(\ref{ring-equation}) is as follows.
Two fluctuations on the same node $\bf r$ that are correlated after collision
at time $t_0$, will be propagated to neighboring nodes ${\bf r}+{\bf c}_i$ 
and ${\bf r}+{\bf c}_j$.  
Due to the collision with other particles at these nodes the correlation will 
be scattered to all directions as described by $(\openone+\Omega)$.
Thus both fluctuations branch into many different paths.
At time $t_0+\tau$ the weight of each path is given by the same factor 
$(1/4)^\tau$.
If two correlated paths end at the same node 
--- a so-called ``ring''-collision --- 
they give rise to on-node precollision correlations,
${\cal C}({\bf 0},t_0+\tau) \sim (1/4)^{2\tau} B(t_0)$,
that change the time evolution of the average occupations according to
Eq.~(\ref{gbeq}).
The actual value of ${\cal C}_{ij}({\bf r},t)$ is a superposition of ``ring''
contributions from source terms at all earlier times, although the dominant
contribution comes from the last few time steps.

The fact that Eq.~(\ref{ring-equation}) is linear in ${\cal C}$ allows us to write
\begin{equation}\label{kernel}
	{\cal C}_{ij}({\bf 0},t) = 
	\sum_{t'=0}^{t-1} \sum_{k,l} K_{ij,kl}(t-t') B_{kl}(t').
\end{equation}
Here $K_{ij,kl}(t-t')$ is a memory kernel which does not depend on any of 
the model parameters --- although it does depend on the system size $L$ --- 
and thus can be constructed once and for all using Eq.~(\ref{ring-equation}).  
This can be done in an efficient manner by exploiting the
rotation and reflection symmetry of $K_{ij,kl}(t-t')$.

After an initial fast decay, the memory function decays algebraically for
large t, as $K_{ij,kl}(t) \propto t^{-\alpha}$ with $\alpha\simeq 1.2$ 
for $L=256$.
When the ring kinetic theory is evaluated numerically, 
on time scales on the order of $10^3$ time steps 
this slow decay leads to the build-up of pair correlations 
that are much larger that what is observed in simulations. 
This excess of correlations would be corrected if we include 
higher order correlations  that are not 
taken into account by the the present form of the ring kinetic theory.  
Therefore it is desirable to cut off the memory kernel for large times, i.e.,
to set $K_{ij,kl}(t)\equiv0$ for $t>t_{\rm cutoff}$.
It is natural to choose the cut-off equal to the time it takes to
travel across the the system: $t_{\rm cutoff}=L$.

By rewriting Eq.~(\ref{ring-equation}) in Fourier representation, it
can be interpreted in terms of modes at different wavevectors ${\bf q}$
(see Ref.~\cite{Bussemaker95a}).
When no reactions occur, the diffusive modes around ${\bf q}={\bf 0}$ and the
(spurious) staggered modes around ${\bf q}=(\pi,\pi)$ play a special role,
since they correspond to conserved densities.  However, in the presence of
reactive collisions there are no conserved densities, and all modes are in
principle equally important in determining the size of the correlations.

In the next section we will compare theoretical predictions 
with the results of computer simulations.  
Numerically, the theory of this section is evaluated as follows.
At time $t=0$ we set $f_i(0)=x_0$ for $1\leq i\leq 4$,
$f_i(0)=y_0$ for $5\leq i\leq 8$, and ${\cal C}_{ij}({\bf r},0)=0$.
To perform a time evolution step from time $t$ to time $t+1$, we first 
use $f(t)$ to calculate the nonlinear Boltzmann operator $I_i[f(t)]$ and the
correction term $I^\prime_{i,kl}[f(t)]$.
Together with the on-node correlations ${\cal C}_{ij}({\bf 0},t)$ we then
use these operators to calculate $f_i(t+1)$ with the help of Eq.~(\ref{gbeq}).
To calculate the evolution of the pair correlation function we use $f_i(t)$
and ${\cal C}_{ij}({\bf 0},t)$
to evaluate the source term $B_{ij}(t)$ in Eq.~(\ref{source}) and then obtain
${\cal C}_{ij}({\bf r},t+1)$ with the help of Eq.~(\ref{ring-equation}).
Iteration of the above procedure yields the set $\{[x(t),y(t)];\ t\geq 0\}$
defining a trajectory in the $x$-$y$ concentration plane.
For large times, either a fixed point or a limit cycle is reached.

\section{Comparison with simulations}\label{sec:compare}

Our simulations were carried out on a $256\times 256$ square lattice. 
The parameters used in the simulations were $k=0.9$ and $c=2$, i.e.,
identical to those used in Ref.~\cite{Dab}.
At $t=0$ the system was prepared in an uncorrelated  homogeneous 
state, with average concentrations $x_0=y_0=0.6$. 
Then we performed the time evolution of the LGA, according to 
section~\ref{sec:model}. 
The initial time steps were discarded, as
the system needs some time to build up the correlations that will 
eventually  produce the shrinking of the limit cycle. 
Once the correlations have been created we record the 
spatially averaged concentration of both species. 
The scale parameter $s$ was varied between $s=2$ and $s=20$. 

In Fig.~\ref{fig1} the dashed line denotes the limit cycle as it is
obtained from the mean-field theory defined by Eqs.~(\ref{Maginu}) and
(\ref{transformation}), assuming that the concentration of both
species are homogeneous and the term $\nabla^2$ can be neglected.
For relatively large values of the time scaling parameter $s$ we expect
mean-field theory to be accurate.  This is confirmed by the simulation data
for $s=20$, shown as a gray band in Fig.~\ref{fig1}, which are reasonably
close to the mean-field prediction.
The width of the gray band corresponds to the fluctuations in the 
spatially averaged concentrations that occur due to the finite system 
size.

When $s$ is decreased correlations become important 
(measurements show that correlations are 
typically 10 times larger for $s=4$ than for $s=20$) 
and the diffusion process is not able to keep the system homogeneous.
As a consequence, different regions in the system become desynchronized 
to a certain degree, and the contribution to the average concentration 
of one region is partially canceled out by out-of-phase contributions 
from other regions. 
This produces a shrinking of the limit cycle in Fig.~\ref{fig1}, as is 
shown by the simulation data for $s=6$ and $s=4$ in Fig.~\ref{fig1}.
The effect is stronger for smaller $s$.

It is clear from Fig.~\ref{fig1} that mean-field theory completely fails
for the smaller values of $s$.
The solid black lines in Fig.~\ref{fig1} represent the limit cycle as it 
is predicted by the ring kinetic theory of section~\ref{sec:theory}.
For the values $s=20$ and $s=6$ shown in Fig.~\ref{fig1} our theory gives an
excellent quantitative prediction of the shrinking of the limit cycle.
For $s=4$ there are deviations due to higher order effects that are not 
taken into account.

Analysis of the ring kinetic theory shows that as $s$ is further
decreased, the limit cycle shrinks continually, until at $s\simeq3$ there 
is an inverse Hopf bifurcation from a limit cycle to a fixed point.
This bifurcation corresponds to a {\em desynchronization transition}, 
where the coherence between different regions is lost completely.
It should be noted that this transition is of a different character than
the Hopf bifurcation that occurs at the mean-field level as a function of 
the model parameters $k$ and $c$.

Let us consider the case $s=2$ in some detail.  Here $s$ is close to the 
smallest possible value (see Ref.~\cite{Dab}) and the fluctuations caused by
the chemical reactions are strongest here. 
Diffusion is not fast enough to keep the system homogeneous except at 
very small scales. The ring kinetic theory for $s=2$ predicts a fixed point
located at $x=y=1/2$.
Simulations for a system of linear size $L=256$ reveal that the average
concentrations fluctuate around the point $x=y=1/2$ in a irregular fashion,
and in a range between 0.47 and 0.53.
In order to assess whether the result of the simulations for $s=2$ corresponds 
to a fixed point or to a limit cycle, we compared numerical simulations
for three different system sizes: $L=32$, $L=256$, and $L=1024$.
The concentration of species $X$ versus time is plotted in 
Fig.~\ref{fig2}. The vertical scale in all three plots is the same.
Clearly, the amplitude of the oscillation decreases with the system size. 
In the $L=32$ system the concentration $x$ oscillates with an amplitude 
$\delta x\simeq0.10$; in the $L=256$ system we have 
$\delta x\simeq0.02$, and in the $L=1024$ system the fluctuations
are very small, $\delta x\simeq0.004$. 
It is reasonable to conclude that for $s=2$ 
the correct solution is a stable fixed 
point, in perfect agreement with our theory. 

In close connection with this last point, we have verified that 
for $s\geq 4$ the limit cycle obtained in the simulations is finite and
stable, and independent of the size of the system up to size $L=1024$.
For $s=3$ our ring kinetic theory predicts a fixed point $x=y=1/2$.
However, simulations are here not conclusive, as systems of intermediate 
size $L=256$ show a limit cycle, but large systems do not reach any 
stationary behavior within available computer time. 
We conclude that the (inverse) Hopf bifurcation from a limit cycle to a
fixed point at the level of the spatially averaged concentrations in a large
enough system must occur between $s=2$ and $s=4$.

The comparison between ring kinetic theory and simulations has so far 
been restricted to the shape of the limit cycle.
Figure~\ref{fig1} however does not give any 
information about the actual time evolution of the concentrations, or
the period of oscillation around the limit cycle. In order to obtain
this information, we have plotted in Fig.~\ref{fig3} the average
concentration of the two species versus time, for both theory and simulations. 
Figure~\ref{fig3}A and B show the concentration of particles of type $X$ 
and $Y$, respectively, for $s=10$. 
Simulation results are indicated by a solid line, while the ring kinetic 
theory is denoted by a dashed line.  
The amplitudes of the oscillation agree quite well, as we 
already knew from Fig.~\ref{fig1}. There is however some deviation between
theoretical and simulated periods, that causes the curves to 
become slightly out of phase.
The difference between both oscillation periods is 
about 3\%.  Figures~\ref{fig3}C and D show similar curves for $s=4$. 
Here the agreement is worse, and the difference in periods is about 11\%. 

Figure~\ref{fig4} shows how the oscillation period --- normalized by dividing
by $s$ --- depends on $s$.
We have plotted the mean-field value of the period as a dashed line;
the ring theory is denoted by circles, and simulation results by triangles. 
It was shown in Fig.~\ref{fig1} that ring kinetic theory predicts the shape 
of the limit cycle quite well down to $s\simeq6$. 
It is therefore somewhat surprising that the mean-field prediction for the 
oscillation period, which is $s$-independent, is better than ring
kinetic theory for all values of $s$.
To resolve this issue it would be necessary to include higher order 
correlation functions in the theoretical description.  
This is clearly beyond the scope of the present paper.
Furthermore, it can be seen that the approach to the mean-field 
period for large $s$ is slow, and even for $s=20$ there is a clear deviation 
of about 1\%.  This effect is probably due to the particular 
choice of the transition rates, that are not able to maintain the 
local diffusive equilibrium even for high $s$ \cite{review-LGA}.

\section{Discussion}\label{sec:discussion}

In this paper we have shown how a theory that takes into account equal-time 
pair correlations provides an excellent 
explanation of the large deviations from mean-field theory observed in 
diffusion-limited chemical reactions as modeled by lattice-gas automata 
(LGA).
Our theory is a straightforward application of the general framework
established in the papers of Ernst and 
coworkers \cite{Bussemaker95a,Ernst+co}.
It is not restricted to the Maginu model, but is applicable to 
any chemical reaction that can be modeled with an LGA.

It is in principle possible to include triplet and higher order correlations
as well.
However, the good agreement between theory and simulations indicates
that the ring theory of section~\ref{sec:theory} captures the
essential physics in a quantitative way.
Although the comparison between theory and simulations reported here
is restricted to the domain of LGA's, we expect that {\em mutatis mutandi}
the general concepts apply equally well to continuous systems.

We have focused on a particular two-species model exhibiting periodic
oscillations of the average concentrations.
Wu and Kapral \cite{Wu} have studied a model with more complicated 
temporal behavior ---
period doubling bifurcations and a transition to a strange attractor, 
as model parameters are changed.
They investigated the consequences of spatial fluctuations by means of
computer simulations.
It is an interesting question whether some of the features observed in
that work can be explained using the theory presented in this paper.
As a final remark we mention that our theory provides a more
microscopic analogue of the Langevin equation method used in 
Ref.~\cite{Weimar} to predict the magnitude of spatial density correlations.

\acknowledgements

It is a great pleasure to dedicate this paper to Matthieu Ernst on the
occasion of his sixtieth birthday.  
Over the last decade, Matthieu has played an important role in 
developing a broad theoretical understanding of the behavior of lattice 
gas automata.
Both authors have greatly enjoyed working with him, and appreciate his 
exceptional sense of responsibility, in particular when it comes to 
training young scientists. 

We thank D. Dab and J. P. Boon for providing us with the
table with the transition rates. R.B. acknowledges financial
support from D.G.I.C.yT. (Spain), project PB94-0265.

\newpage

\begin{figure}
\caption{Average concentrations $x$ and $y$ in the Maginu model for several 
values of the time scaling parameter: $s=4, 6$, and $20$. 
The outer dashed line corresponds to mean-field theory, given by
Eqs.~(\protect\ref{Maginu}) and (\protect\ref{transformation}). 
Solid lines correspond to the ring kinetic theory of 
section~\protect\ref{sec:theory}.
The gray bands denote the result of computer simulations performed on 
square lattices of size $256\times 256$;
their width corresponds to the size of the fluctuations from cycle 
to cycle.}
\label{fig1}
\end{figure}

\begin{figure}
\caption{Average concentration $x$ of species $X$ versus time $t$ for the
Maginu model at $s=2$, and system size $32\times32$, $256\times256$,
and $1024\times1024$, respectively. 
The bigger the system, the smaller the fluctuations.}
\label{fig2}
\end{figure}

\begin{figure}
\caption{Concentration of species $X$ and $Y$ versus time $t$ for
$s=10$ (figures A and B), and for $s=4$ (figures C and D). Solid lines
are the results of the computer simulations (in systems of size
$256\times256$), while the dashed lines correspond to ring kinetic theory.}
\label{fig3}
\end{figure}

\begin{figure}
\caption{Oscillation period as a function of $s$. The mean 
field value is indicated by a dashed line. Circles denote
the ring kinetic theory prediction of section~\protect\ref{sec:theory}.
Triangles are  the simulation values. 
Mean-field theory is here in better agreement with simulations 
than ring kinetic theory.} 
\label{fig4}
\end{figure}

\newpage
\vspace*{2cm}
\[ 
\psfig{file=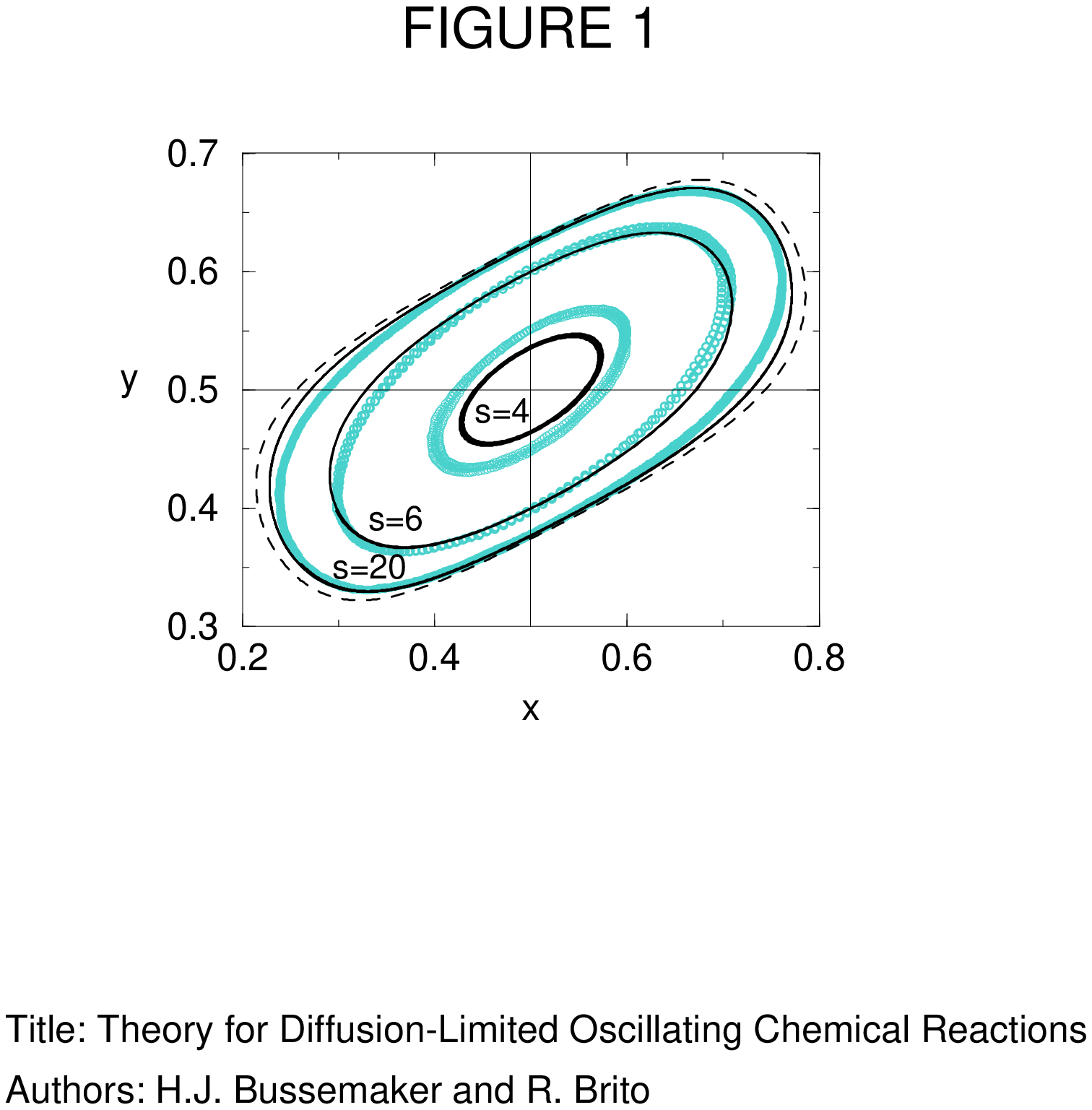} 
\]

\newpage
\vspace*{2cm}
\[ 
\psfig{file=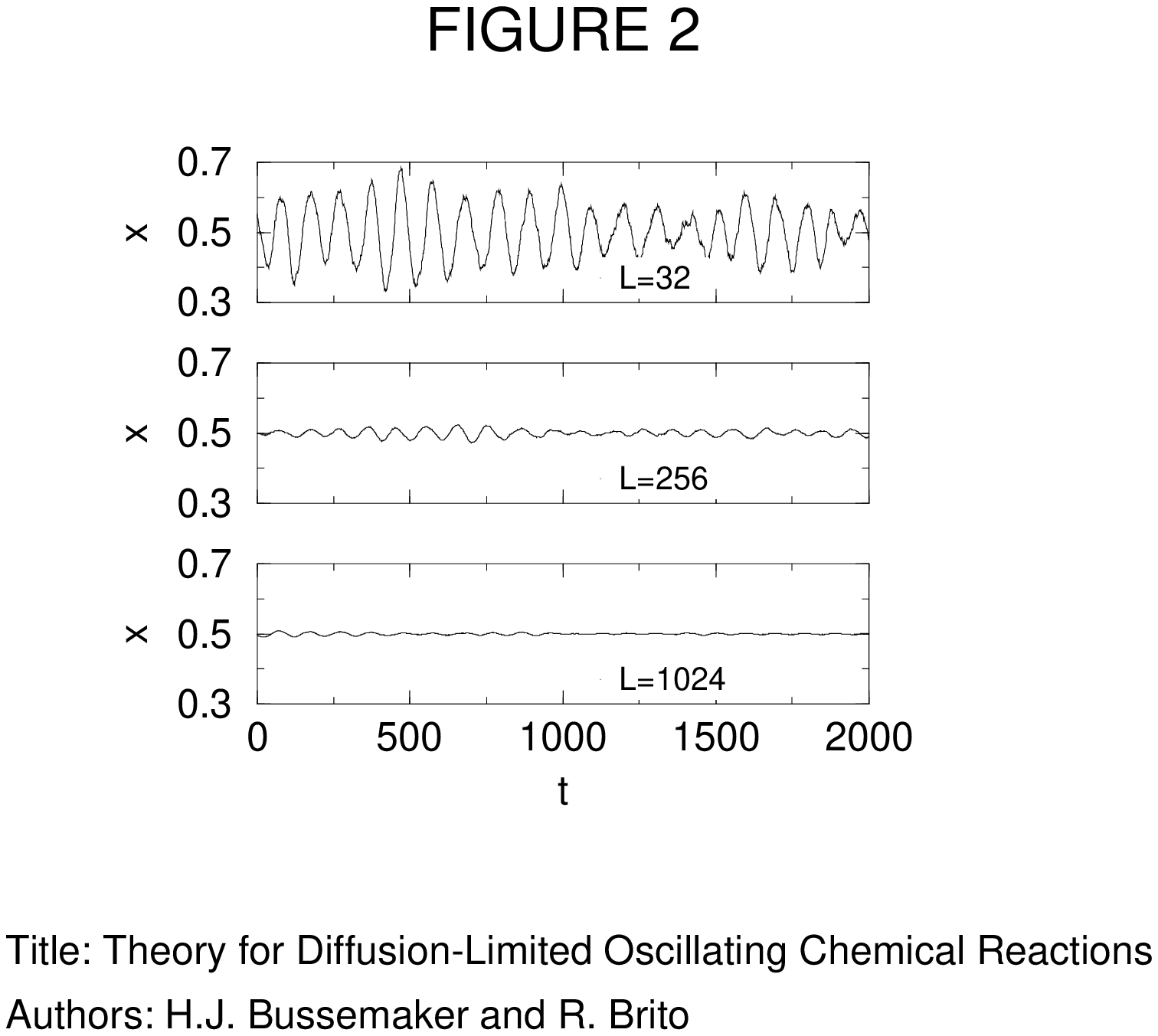} 
\]

\newpage
\vspace*{2cm}
\[ 
\psfig{file=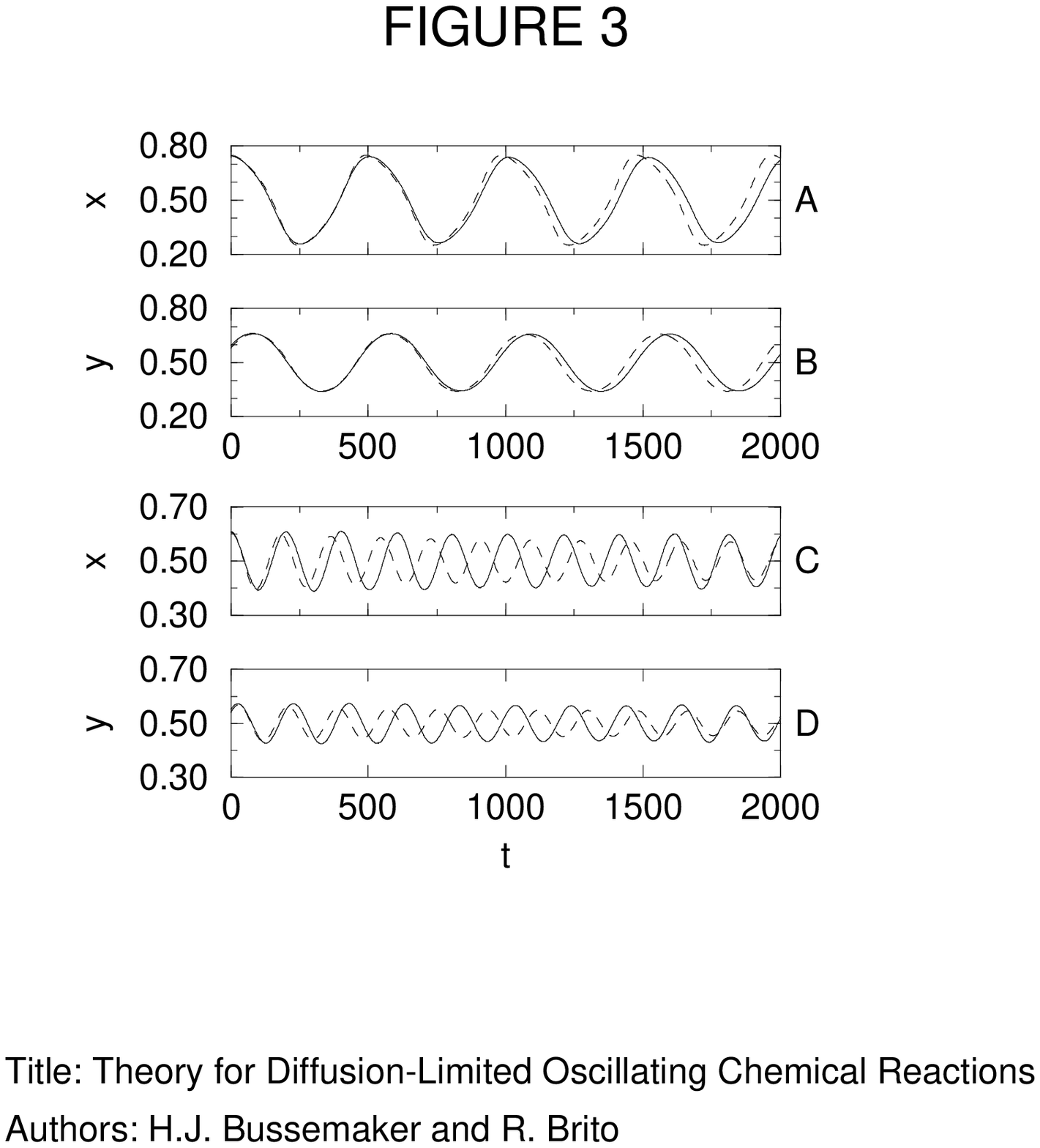} 
\]

\newpage
\vspace*{2cm}
\[ 
\psfig{file=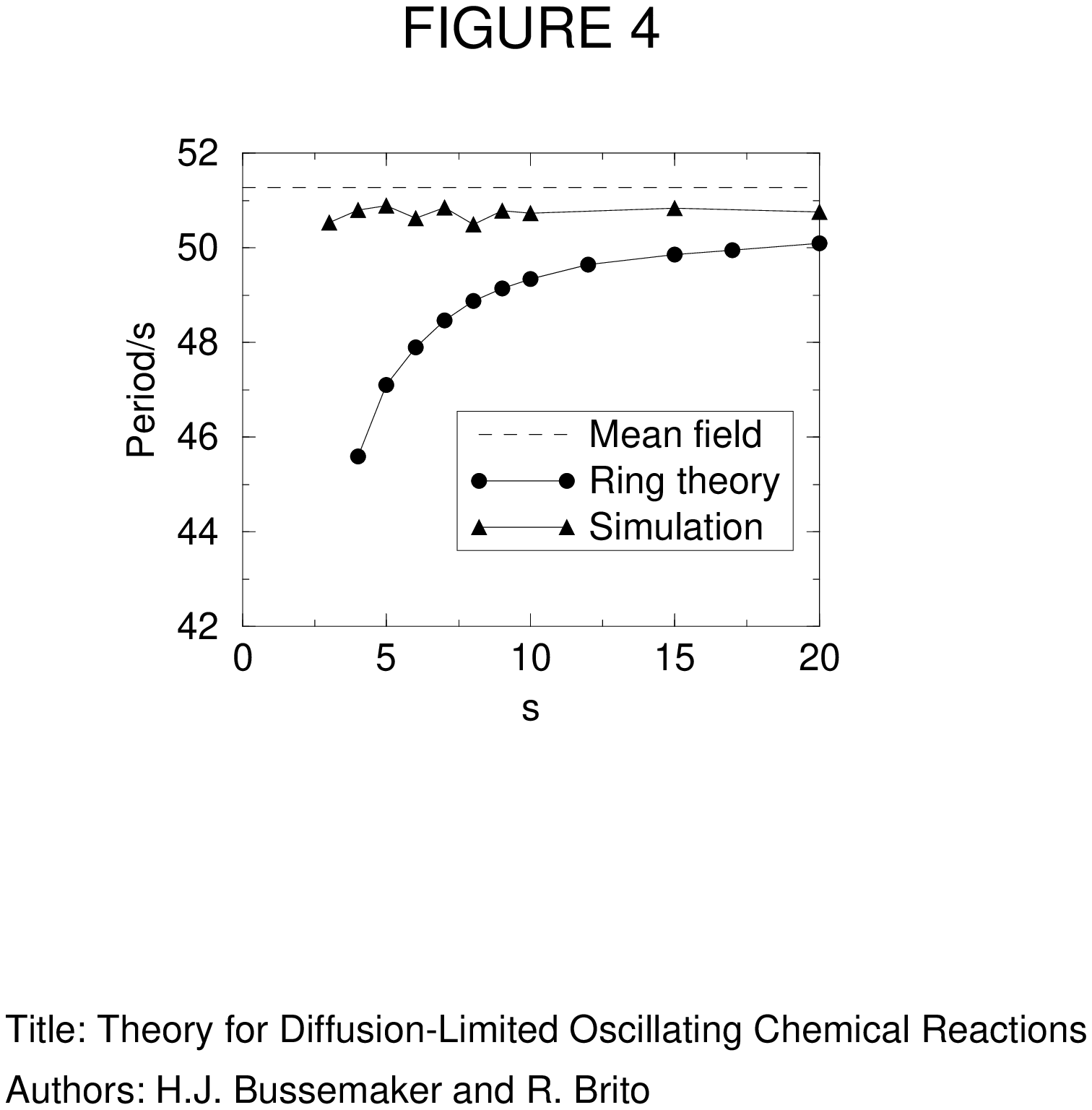} 
\]

\end{document}